\documentclass[11pt,a4paper]{article}
\usepackage{mathtools,cancel}

\usepackage{soul,xcolor}
\usepackage{xcolor}
\usepackage{multirow}
\usepackage{pstricks}
\usepackage{dcolumn}
\usepackage{bm}
\usepackage{latexsym}
\usepackage{dcolumn}
\usepackage[utf8x]{inputenc} 
\usepackage{amsmath}
\usepackage{amsfonts,amssymb}
\usepackage{graphicx,epsfig}
\usepackage{color}
\usepackage{psfrag}
\usepackage{amsthm}
\usepackage{ulem} % compatiple with \sout
\usepackage{soul} % compatiple with \st
\usepackage{flushend}
\usepackage{titlesec}
\usepackage{slashed}
\usepackage{authblk}
\usepackage[font=footnotesize]{caption}
\usepackage[hmarginratio=1:1,top=32mm,columnsep=60pt]{geometry}
\newcommand{\bea}{\begin{eqnarray}}
\newcommand{\eea}{\end{eqnarray}}
\newcommand{\be}{\begin{equation}}
\newcommand{\ee}{\end{equation}}
\newcommand{\ba}{\begin{array}}
\newcommand{\ea}{\end{array}}

\def \lb{\Lambda_b}
\def \lc{\Lambda_c}
\def\beq{\begin{equation}}
\def\eeq{\end{equation}}
\def\bea{\begin{eqnarray}}
\def\eea{\end{eqnarray}}

\def\bra{\langle}
\def\ket{\rangle}

\begin{document}
%%%%%%%%%%%%%%%%%%%%%%%%%%%%%%%%%%%%%%%%%%%%%%%%%%%%%%%%%%% Title, Authors and Citations %%%%%%%%%%%%%%%%%%%%%%%%%%%%%%%%%%%%%%%%%%%%%%%%%%%%%%%%%%%%%%%%%%%

\unitlength = 1mm
\begin{flushleft}
PHENO-2023
%[10mm]
\end{flushleft}

\begin{center}
\bigskip {\Large  \bf C-Code Reader of Form Factors}
\\[8mm]
%Bradley Smith  $^{\dag}$ 
%\footnote{E-mail:
%\texttt{bs5797@cs.ship.edu}} 
Ahmed Rashed $^{\dag}$  
\footnote{E-mail:
\texttt{amrashed@ship.edu}} 
%and Chen Huo $^{\ddag}$  
%\footnote{E-mail:
%\texttt{chuo@ship.edu}} 
\\[3mm]
\end{center}

\begin{center}
~~~{\it $^{\dag}$ Department  of Physics,} 
{\it Shippensburg University of Pennsylvania,}\\
~~~{\it  Franklin Science Center,}
{\it 1871 Old Main Drive, Pennsylvania, 17257, USA}\\
\end{center}

%\newpage

\begin{center} 
\bigskip (\today) \vskip0.5cm {\Large Abstract\\} \vskip3truemm
\parbox[t]{\textwidth}  
{
The process $\Lambda_b \rightarrow \Lambda_c \ell^- \bar{\nu}_\ell$ serves as a tool for exploring new physics, with contributions from scalar, vector, and tensor hadronic currents in various models. These form factors are derived from the quark model or lattice QCD. This work introduces a C-code for efficiently reading lattice QCD form factors for these currents, significantly outperforming a previous Mathematica-based approach, with a speed improvement of over 64 times per data point. The code, available on GitHub $(https://github.com/darkfiresmith96/Lattice_QCD)$, also features a web interface for user inputs.}
\end{center}

\thispagestyle{empty} \newpage \setcounter{page}{1}
% Decrease texheight (for preprint numbers) again
%\textheight 23.0 true cm
\baselineskip=14pt

\section{Introduction}

The process $\Lambda_b \to \Lambda_c \ell^- \bar{\nu}_\ell$ is a valuable probe for physics beyond the Standard Model (SM), offering insights into new physics (NP) contributions from scalar, vector, and tensor currents. It tests the limits of the Non-Relativistic Quark (NRQ) model \cite{Chakraverty:1996sb,Albertus:2004wj} and helps investigate anomalies such as $R(D^{(*)})$ and $R(K^{(*)})$ \cite{Datta:2017aue, Shivashankara:2015cta}. These anomalies suggest potential lepton universality violations in $\bar{B} \to D^{(*)}\ell \nu_\ell$ \cite{Lees:2013uzd,Less:2013uzd2,Bhattacharya:2014wla,Bhattacharya:2016mcc} and $B \to K^{(*)} \ell^+ \ell^-$ \cite{Aaij:2014ora,Bhattacharya:2014wla,Bhattacharya:2016mcc}, possibly linked to new heavy \cite{Alok:2017sui,Datta:2019zca} or light states \cite{Datta:2017pfz,Datta:2017ezo}. 

Additionally, the decay rates of $\Lambda_b \to p \ell^- \bar{\nu}_\ell$ and $\Lambda_b \to \Lambda_c \ell^- \bar{\nu}_\ell$ are crucial for determining CKM matrix elements $V_{ub}$ and $V_{cb}$ \cite{Datta:1995mv}, necessitating precise form factor calculations. The form factors can be derived from quark models \cite{Cardarelli:1997sx,Dosch:1997zx,Huang:1998rq,MarquesdeCarvalho:1999bqs,Huang:2004vf,Pervin:2005ve,Ke:2007tg,Wang:2009hra,Azizi:2009wn,Khodjamirian:2011jp,Gutsche:2014zna} or lattice QCD \cite{Detmold:2015aaa, Datta:2017aue}, which parameterizes the hadronic transition $\Lambda_b \to \Lambda_c$ via scalar, vector, and tensor form factors.

The LHCb Collaboration recently measured the ratio of decay rates for $\Lambda_b \to p \ell^- \bar{\nu}_\ell$ and $\Lambda_b \to \Lambda_c \ell^- \bar{\nu}_\ell$ \cite{Aaij:2015bfa}. Lattice QCD calculations have been employed to predict differential and integrated decay rates for these processes. In Ref. \cite{Datta:2017aue}, lattice QCD form factors for the scalar, vector, and tensor currents were used, building on previous work with 2+1 flavors of dynamical domain-wall fermions \cite{Detmold:2015aaa}. Tensor form factors were extracted using ratios defined in Ref. \cite{Detmold:2016pkz}, with uncertainties evaluated through nominal and higher-order fits. Supplemental data files provided covariance matrices for all ten $\Lambda_b \to \Lambda_c$ form factors. Additionally, a Mathematica code was developed to read these form factors for analysis.

\section{Lattice Form Factor}

In the presence of new physics (NP), the effective Hamiltonian for the quark-level transition \( b \to c\tau^-\bar{\nu}_\tau \) is given as follows \cite{Chen:2005gr,Bhattacharya:2011qm,Datta:2012qk}:

\[
\begin{aligned}
\mathcal{H}_{\text{eff}} &= \frac{G_F V_{cb}}{\sqrt{2}} \Big\{ \Big[\bar{c} \gamma_\mu (1-\gamma_5) b + g_L \bar{c} \gamma_\mu (1-\gamma_5) b + g_R \bar{c} \gamma_\mu (1+\gamma_5) b\Big] \bar{\tau} \gamma^\mu (1-\gamma_5) \nu_\tau \\
&\quad + \Big[g_S \bar{c} b + g_P \bar{c} \gamma_5 b\Big] \bar{\tau} (1-\gamma_5) \nu_\tau + \Big[g_T \bar{c} \sigma^{\mu \nu} (1-\gamma_5) b\Big] \bar{\tau} \sigma_{\mu \nu} (1-\gamma_5) \nu_\tau + \text{h.c.} \Big\},
\end{aligned}
\]
where \( G_F \) is the Fermi constant, \( V_{cb} \) is the CKM matrix element, and \( \sigma_{\mu \nu} = i[\gamma_\mu, \gamma_\nu]/2 \). The Hamiltonian is evaluated at the energy scale \( m_b \).

The hadronic helicity amplitudes for the transition \( \Lambda_b(p_{\Lambda_b}) \to \Lambda_c(p_{\Lambda_c}) \) are parameterized in terms of scalar, vector/axial-vector, and tensor form factors as:

\textbf{Scalar and Pseudoscalar Contributions}
\[
\begin{aligned}
H^{SP}_{\lambda_{\Lambda_c},\lambda=0} &= H^S_{\lambda_{\Lambda_c},\lambda=0} + H^P_{\lambda_{\Lambda_c},\lambda=0}, \\
H^S_{\lambda_{\Lambda_c},\lambda=0} &= g_S \bra{\Lambda_c} \bar{c} b \ket{\Lambda_b}, \\
H^P_{\lambda_{\Lambda_c},\lambda=0} &= g_P \bra{\Lambda_c} \bar{c} \gamma_5 b \ket{\Lambda_b}.
\end{aligned}
\]

\textbf{Vector and Axial-Vector Contributions}
\[
\begin{aligned}
H^{VA}_{\lambda_{\Lambda_c},\lambda} &= H^V_{\lambda_{\Lambda_c},\lambda} - H^A_{\lambda_{\Lambda_c},\lambda}, \\
H^V_{\lambda_{\Lambda_c},\lambda} &= (1+g_L+g_R) \epsilon^{*\mu}(\lambda) \bra{\Lambda_c} \bar{c} \gamma_\mu b \ket{\Lambda_b}, \\
H^A_{\lambda_{\Lambda_c},\lambda} &= (1+g_L-g_R) \epsilon^{*\mu}(\lambda) \bra{\Lambda_c} \bar{c} \gamma_\mu \gamma_5 b \ket{\Lambda_b}.
\end{aligned}
\]

\textbf{Tensor Contributions}
\[
\begin{aligned}
H^{(T)}_{\lambda_{\Lambda_c},\lambda ,\lambda^{\prime}} &= H^{(T1)}_{\lambda_{\Lambda_c},\lambda ,\lambda^{\prime}} - H^{(T2)}_{\lambda_{\Lambda_c},\lambda ,\lambda^{\prime}}, \\
H^{(T1)}_{\lambda_{\Lambda_c},\lambda ,\lambda^{\prime}} &= g_T \epsilon^{*\mu}(\lambda) \epsilon^{*\nu}(\lambda^{\prime}) \bra{\Lambda_c} \bar{c} i \sigma_{\mu \nu} b \ket{\Lambda_b}, \\
H^{(T2)}_{\lambda_{\Lambda_c},\lambda ,\lambda^{\prime}} &= g_T \epsilon^{*\mu}(\lambda) \epsilon^{*\nu}(\lambda^{\prime}) \bra{\Lambda_c} \bar{c} i \sigma_{\mu \nu} \gamma_5 b \ket{\Lambda_b}.
\end{aligned}
\]

\textbf{Scalar and Pseudoscalar Matrix Elements}
\bea
\nonumber \bra{\lc}\bar{c} b\ket{\lb} &=& \frac{q_\mu}{m_b-m_c}\bra{\lc}\bar{c}\gamma^\mu b\ket{\lb} \\
&=& F_0(q^2)  \frac{m_{\lb} - m_{\lc}}{m_b-m_c} \bar{u}_{\lc}u_{\lb}, \\
\nonumber \bra{\lc}\bar{c}\gamma_5 b\ket{\lb} &=& \frac{q_\mu}{m_b+m_c}\bra{\lc}\bar{c}\gamma^\mu\gamma_5 b\ket{\lb} \\
&=& G_0(q^2)  \frac{m_{\lb} + m_{\lc}}{m_b+m_c} \bar{u}_{\lc}\gamma_5 u_{\lb}.
\eea

\textbf{Vector and Axial-Vector Matrix Elements}
\[
\begin{aligned}
\bra{\Lambda_c} \bar{c} \gamma^\mu b \ket{\Lambda_b} &= \bar{u}_{\Lambda_c} \Big[ F_0 (q^2)(m_{\Lambda_b} - m_{\Lambda_c})\frac{q^\mu}{q^2} \\
&\quad + F_+ (q^2)\frac{m_{\Lambda_b} + m_{\Lambda_c}}{Q_+}(p_{\Lambda_b}^{\mu} +p_{\Lambda_c}^{\mu}-(m_{\Lambda_b}^2 - m_{\Lambda_c}^2)\frac{q^\mu}{q^2}) \\
&\quad + F_\perp (q^2)(\gamma^\mu - \frac{2m_{\Lambda_c}}{Q_+}p_{\Lambda_b}^{\mu} - \frac{2m_{\Lambda_b}}{Q_+}p_{\Lambda_c}^{\mu})\Big]u_{\Lambda_b},
\end{aligned}
\]
\[
\begin{aligned}
\bra{\Lambda_c} \bar{c} \gamma^\mu \gamma_5 b \ket{\Lambda_b} &= -\bar{u}_{\Lambda_c} \gamma_5 \Big[ G_0 (q^2)(m_{\Lambda_b} + m_{\Lambda_c})\frac{q^\mu}{q^2} \\
&\quad + G_+ (q^2)\frac{m_{\Lambda_b} - m_{\Lambda_c}}{Q_-}(p_{\Lambda_b}^{\mu} +p_{\Lambda_c}^{\mu}-(m_{\Lambda_b}^2 - m_{\Lambda_c}^2)\frac{q^\mu}{q^2}) \\
&\quad + G_\perp (q^2)(\gamma^\mu + \frac{2m_{\Lambda_c}}{Q_-}p_{\Lambda_b}^{\mu} - \frac{2m_{\Lambda_b}}{Q_-}p_{\Lambda_c}^{\mu})\Big]u_{\Lambda_b}.
\end{aligned}
\]

\textbf{Tensor Matrix Elements}
\bea
&&\bra{\lc}\bar{c}i\sigma^{\mu\nu} b\ket{\lb}=\bar{u}_{\lc}\Big[2h_+(q^2)\frac{p_{\lb}^\mu p_{\lc}^{ \nu}-p_{\lb}^\nu p_{\lc}^{\mu}}{Q_+} \nonumber\\
&&+h_\perp (q^2)\Big(\frac{m_{\lb}+m_{\lc}}{q^2}(q^\mu \gamma^\nu -q^\nu \gamma^\mu)-2(\frac{1}{q^2}+\frac{1}{Q_+})(p_{\lb}^\mu p_{\lc}^{\nu}-p_{\lb}^\nu p_{\lc}^{\mu}) \Big) \nonumber\\
&&+\widetilde{h}_+ (q^2)\Big(i\sigma^{\mu \nu}-\frac{2}{Q_-}(m_{\lb}(p_{\lc}^{\mu}\gamma^\nu -p_{\lc}^{\nu}\gamma^\mu)\nonumber\\
&&-m_{\lc}(p_{\lb}^\mu \gamma^\nu -p_{\lb}^\nu \gamma^\mu)+p_{\lb}^\mu p_{\lc}^{\nu}-p_{\lb}^\nu p_{\lc}^{\mu}) \Big) \nonumber\\
&&+\widetilde{h}_\perp(q^2) \frac{m_{\lb}-m_{\lc}}{q^2 Q_-}\Big((m_{\lb}^2-m_{\lc}^2-q^2)(\gamma^\mu p_{\lb}^\nu - \gamma^\nu p_{\lb}^\mu)\nonumber\\
&&-(m_{\lb}^2-m_{\lc}^2+q^2)(\gamma^\mu p_{\lc}^{\nu}-\gamma^\nu p_{\lc}^{\mu})+2(m_{\lb}-m_{\lc})(p_{\lb}^\mu p_{\lc}^{\nu}-p_{\lb}^\nu p_{\lc}^{\mu}) \Big)
\Big]u_{\lb}. \nonumber \\ \label{eq:TFF}
\eea

The covariance matrices for all ten form factors, along with the differential decay rate for \( \Lambda_b \to \Lambda_c \ell^- \bar{\nu}_\ell \), are provided in supplemental files. Differential decay rates will be computed for the SM and NP scenarios using a \( \text{C} \)-code for validation against Mathematica results.

\section{Program Overview}

This paper introduces a C-language code for reading scalar, vector, and tensor lattice QCD form factors of the transition \(\Lambda_b \to \Lambda_c\), generating data points for the differential decay rate of the process \(\Lambda_b \to \Lambda_c \tau \bar{\nu}_\tau\). The code, converted from the Mathematica code used in Ref.~\cite{Datta:2017aue}, is available on GitHub $(https://github.com/darkfiresmith96/Lattice_QCD)$ under the GNU General Public License 3 (GPL). The primary code is organized into main.c and supplementary files (Constant.c, Constant.h, Calculate.c, Calculate.h, File$\_$Manipulate.c, File$\_$Manipulate.h), along with data files like covariance, results, HO$\_$covariance, and HO$\_$results.

Verification against the Mathematica code showed that the C-code offers significant efficiency improvements, with a runtime ratio of 1:64.2 per data point, due to its lower-level implementation. The modularity and speed of C also enhance memory and resource control compared to Mathematica.

The code development involved three main steps:
1. \textbf{File Reading}: Four custom read-file functions were created to handle diverse input data structures, which were reorganized for processing.
2. \textbf{Calculations}: Mapping data flow from the Mathematica code required recreating its functions in C and identifying computational patterns, such as those for nominal and higher-order form factors. Complex decay rate formulas were broken down for implementation.
3. \textbf{Data Output}: Results were tabulated, including \(q^2\), differential decay rates, and associated errors, and exported to text files for graphing, such as plotting the differential decay rate versus \(q^2\). 

This work highlights the advantages of C for scientific computing, particularly in efficiency and flexibility, while maintaining consistency with previous results from lattice QCD computations.

\section{Results}

\begin{figure}[h]
 \centering
 \includegraphics{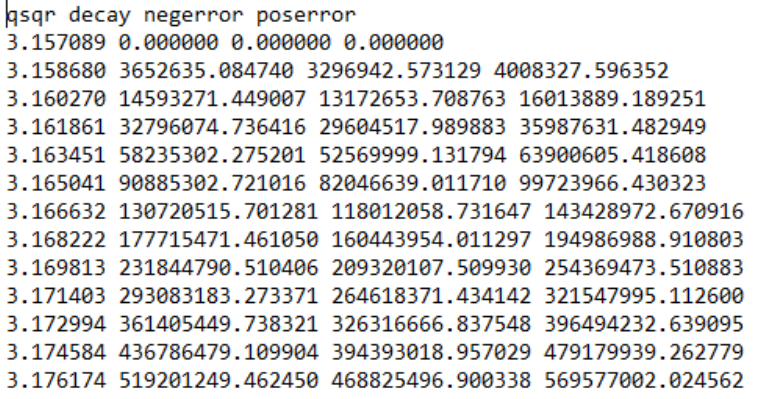}
 \caption{Example of data from code}
 \label{figure1}
\end{figure}

\begin{figure}[!h]
 \centering
 \includegraphics[scale=0.55]{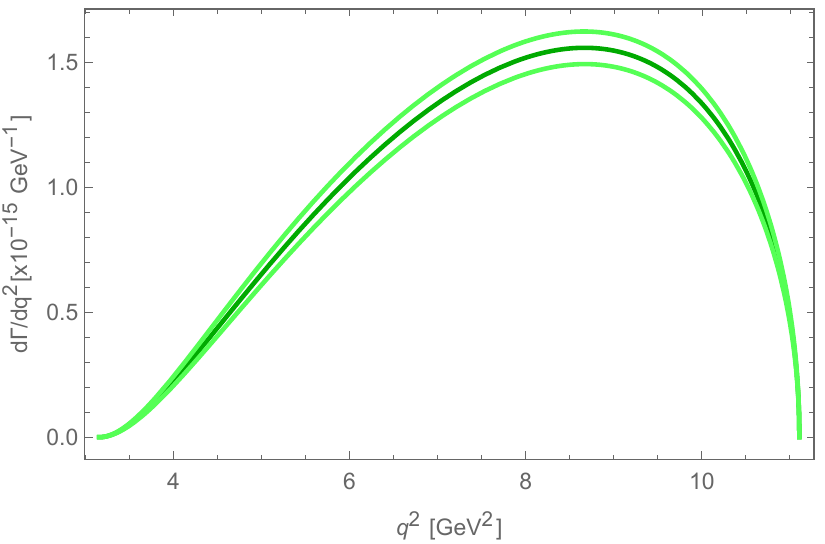}~~~
 \includegraphics[scale=0.55]{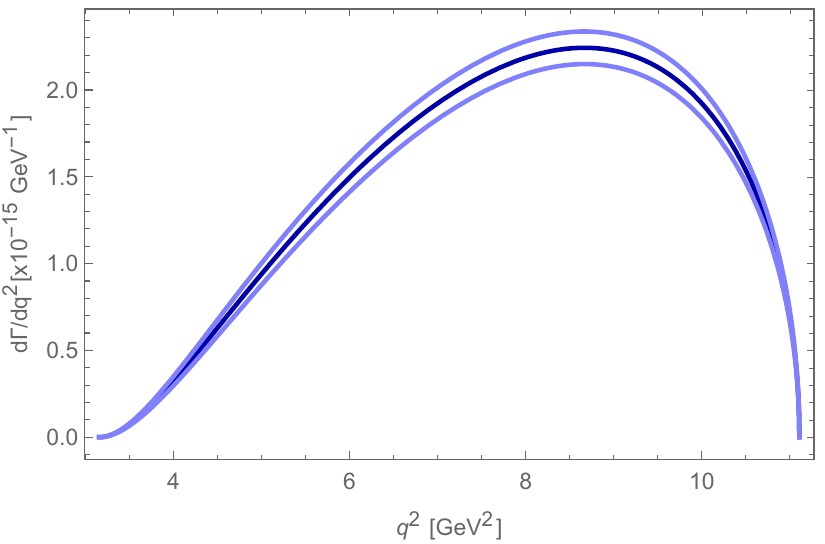}~~~
 \caption{The differential cross section distribution from the Mathematica code. The left panel is the Standard  Model results and the right panel is the new physics results at $g_s= g_p= g_R= g_T=0$ and $g_L=-2.2$.}
 \label{figure2}
\end{figure}

\begin{figure}[!h]
% \centering
\hspace*{-1cm} 
 \includegraphics[scale=0.33]{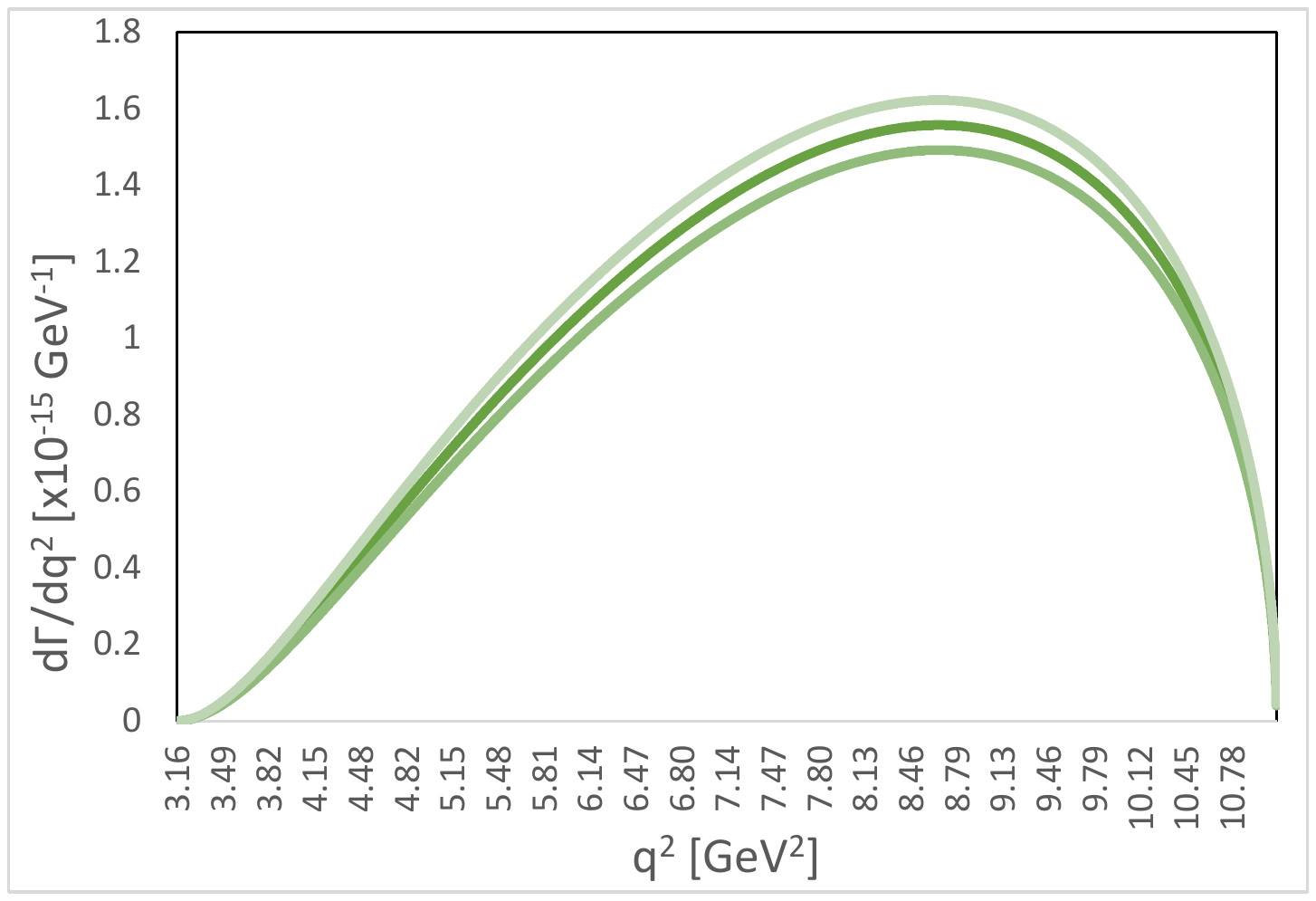}~
 \includegraphics[scale=0.33]{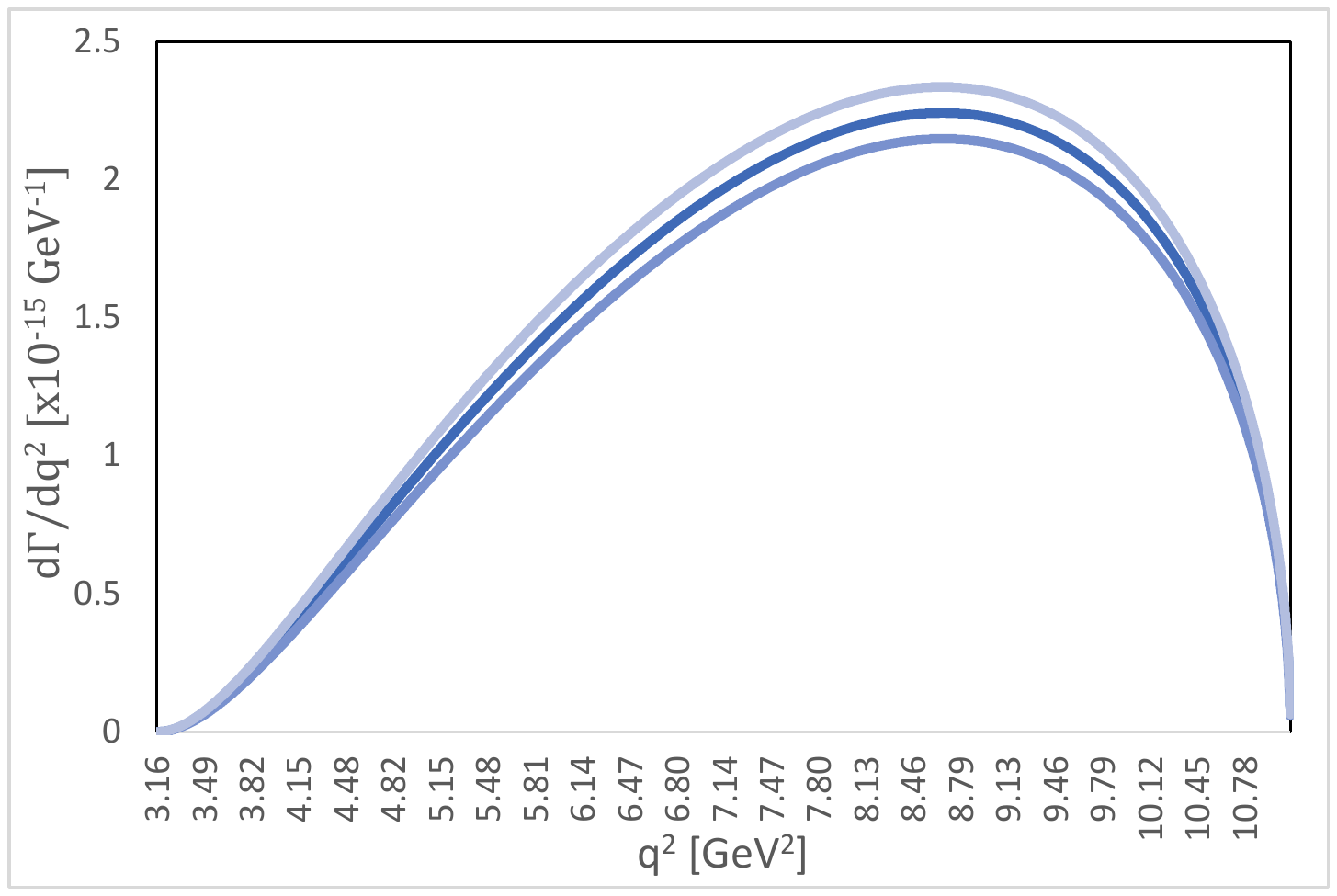}~
 \caption{The differential cross section distribution from the c-code. The left panel is the Standard  Model results and the right panel is the new physics results at $g_s= g_p= g_R= g_T=0$ and $g_L=-2.2$.}
 \label{figure3}
\end{figure}

Once the results are stored in a file, they can be imported into graphing software to create plots of the decay rate and its associated error. The format of the data file is illustrated in Fig.~\ref{figure1}. The resulting graphs of the differential distribution, generated using both the Mathematica and C codes, are shown in Figs.~(\ref{figure2}, \ref{figure3}). These graphs demonstrate that the C-code results align closely with the Mathematica results and are consistent with Ref.~\cite{Datta:2017aue}, confirming the correctness of the C implementation.

The speed comparison between the C and Mathematica codes is summarized in Fig.~4 and Table~1. Both codes were tested with 10, 100, 1000, and 10,000 data points over five trials per case. The wall clock time for each trial was recorded and averaged. The resulting graph of wall clock time versus the number of data points is linear for both codes. Based on the slopes of these graphs, the C-code is determined to be approximately 64.2 times faster than the Mathematica code per data point.

\pagebreak

%\begin{figure}[!h]
% \centering
% \includegraphics[scale=0.9]{C-SM.png}~
% \includegraphics[scale=0.9]{C-NP.png}~
% \caption{The differential cross section distribution from the c-code. The left panel is the Standard  Model results and the right panel is the new physics results at $g_s= g_p= g_R= g_T=0$ and $g_L=-2.2$.}
% \label{figure3}
%\end{figure}

\begin{figure}[!h]
 \centering
 \includegraphics[scale=0.43]{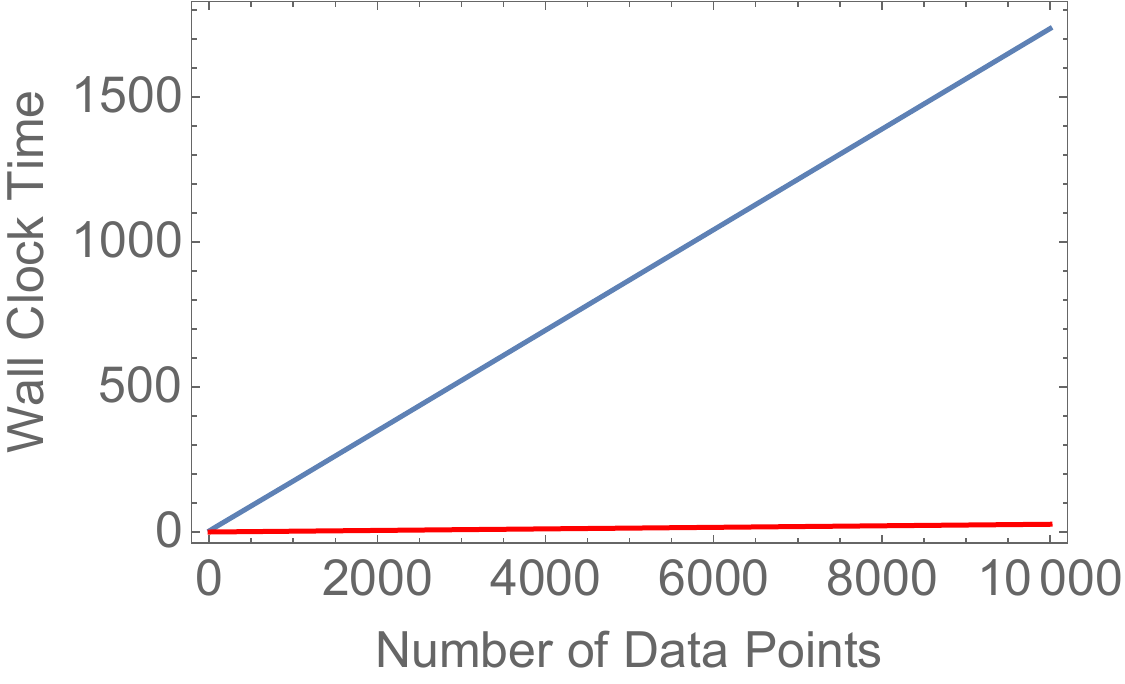}~~~
  \includegraphics[scale=0.43]{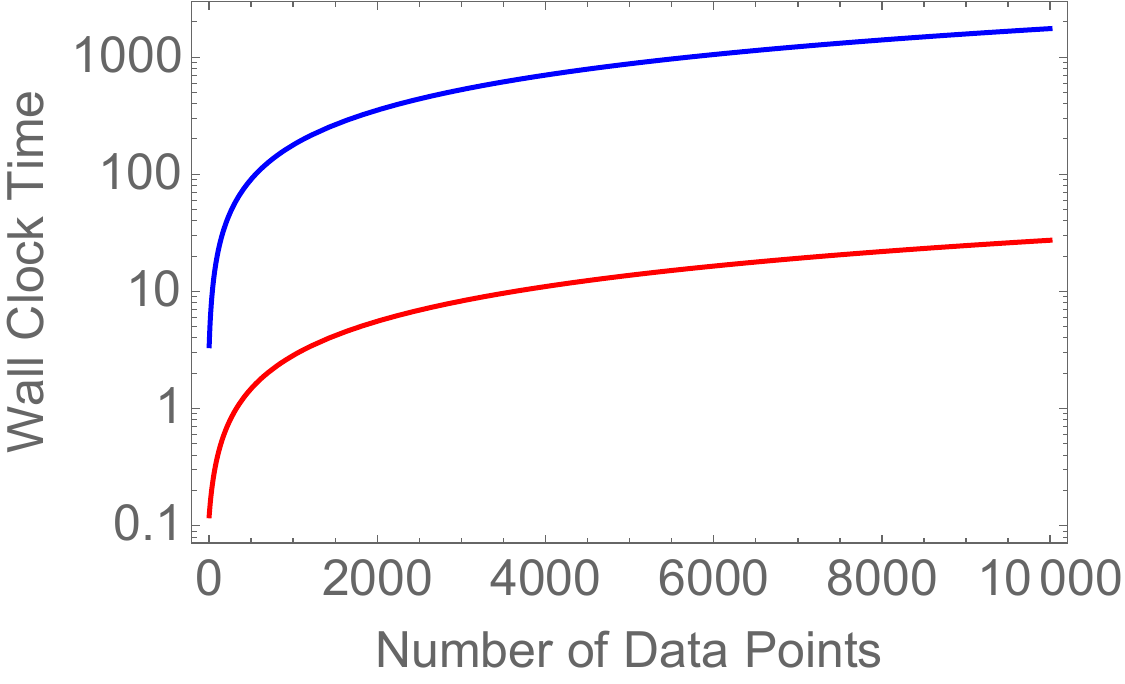}~~~
 \caption{Results of of the speed test. This shows that the c-code (red) performs faster than the Mathematica code (blue). Left and right panels are the linear and log plots, consequently, of the the wall clock time versus the number of data points.}
 \label{figure4}
\end{figure}

\begin{table}[!h]
 \begin{tabular}{||c | c | c ||} 
 \hline
 Number of points & Average Time of Mathematica Code & Average Time of C Code   \\ [0.5ex] 
 \hline\hline
 10 & 5.4264  &0.1076   \\ 
 \hline
 100  & 21.2736 & 0.3652  \\
 \hline
 1000 & 175.7828 & 2.8656  \\
 \hline
  10000 & 1736.1958 & 26.8708   \\ 
  \hline
\end{tabular}
\caption{Results of of the speed test for the c-code and the Mathematica code. }
\label{Table1}
\end{table}

\pagebreak

%\section{Future of Code}
%In the future of the code, a Graphical User Interface (GUI) will be made for easier use. The GUI will include the ability to select the desired file from the computer's file explorer, the ability to select the values of all variables, and the ability to change the number of data points created.  These functions will expand the capability of the code without having to go in and edit it directly, as well as create ease of use.
%Another plan for the code is to expand its scope to create data files for the higher order decay rates, for it only gives data for the nominal, as well as other desirable elements the code calculates.  It will also be expanded to solve similar problems.

\section{Web Application with Graphical Interface}
\begin{figure}[h]
 \centering
 \includegraphics[width=0.8\textwidth]{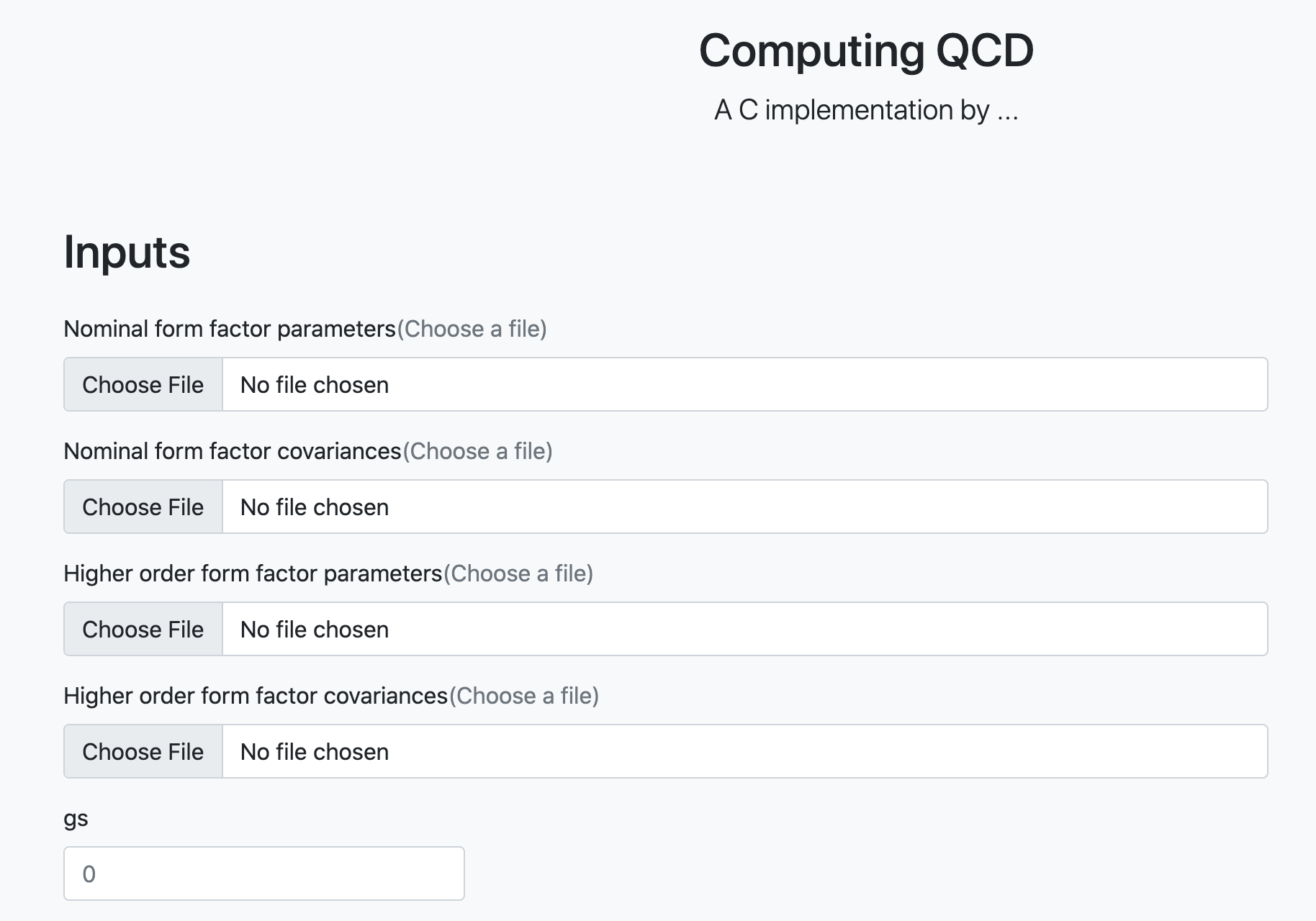}
 \caption{Web interface for user inputs}
 \label{fig:web-interface}
\end{figure}

To expand accessibility beyond users familiar with compiling and running C code on a Linux command line, a prototype web application was developed using the Node.js framework. This web application allows users to upload the necessary data documents and input parameters via a graphical interface, as shown in Fig.~\ref{fig:web-interface}. The application then executes the C program on the server and delivers the computed raw data to the user through the webpage. Hosted on a "free-tier" Heroku Cloud instance, the system utilizes Docker for easy scaling to more powerful computational resources when needed.

\begin{figure}[h]
 \centering
 \includegraphics[width=0.8\textwidth]{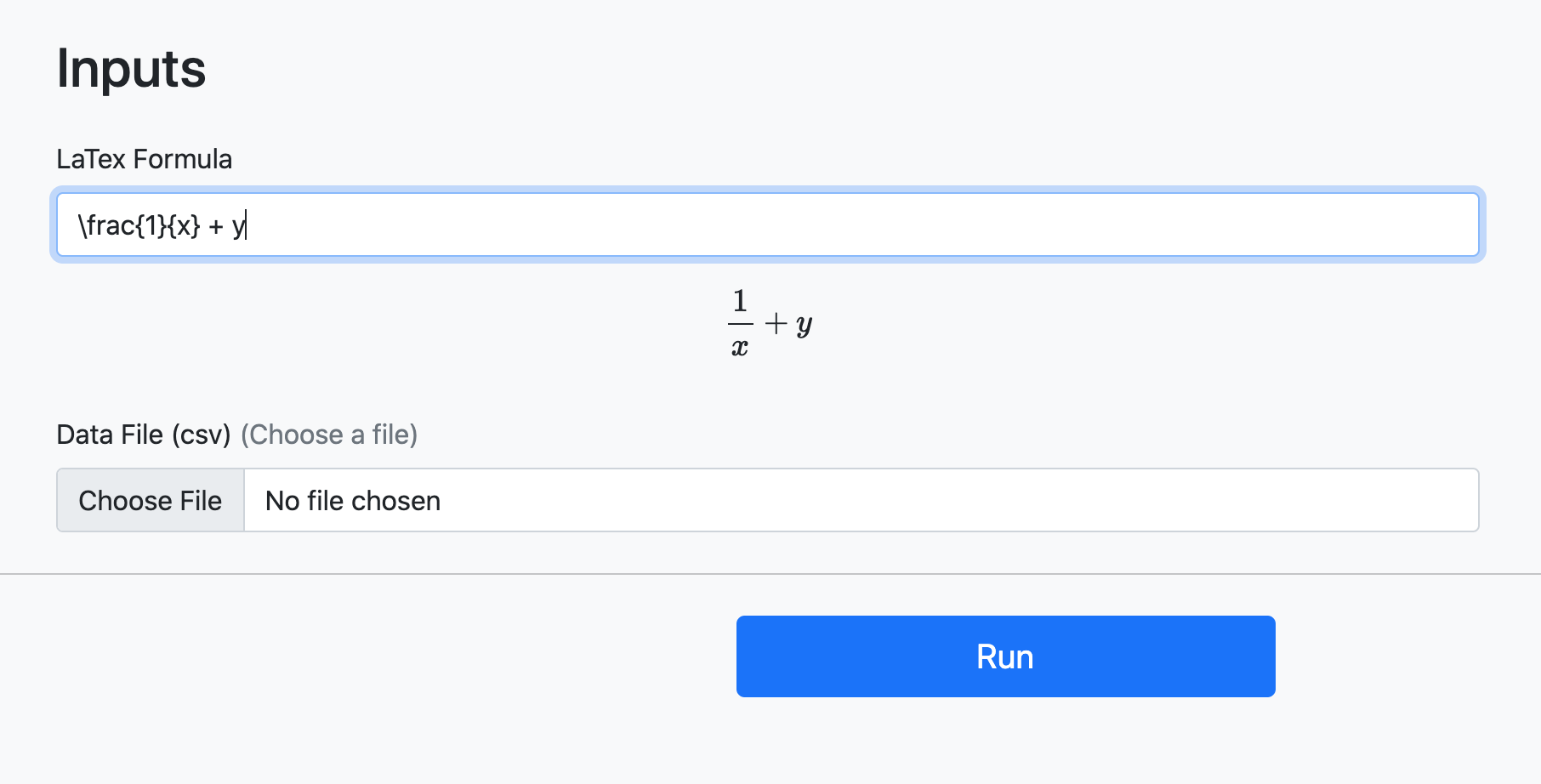}
 \caption{Web application which can parse user's expression}
 \label{fig:web-future}
\end{figure}

Future enhancements include a generalized interface that accepts user-defined mathematical expressions in LaTeX format. Users will upload compatible data files, as illustrated in Fig.~\ref{fig:web-future}. The application parses the input expression, generates JavaScript code to compute the results, and delivers them based on the provided data files (e.g., a spreadsheet with "x" and "y" columns). Future versions will generate and compile C code for improved efficiency. This evolution of the web application demonstrates its potential for broader user adoption and versatility.

\section{Conclusion}

The hadronic transition \(\Lambda_b \to \Lambda_c\) is described using scalar, vector, and tensor QCD lattice form factors. While previous work utilized a Mathematica code for reading form factor data files, this paper presents a C-code alternative that generates decay rate data points for the process \(\Lambda_b \to \Lambda_c \,\tau^- \bar{\nu}_\tau\). Verification against the Mathematica code shows that both provide identical results. Performance comparisons reveal that the C-code is significantly more efficient, with a wall clock time ratio of 1:64.2 per data point compared to the Mathematica code. Additionally, the C-code is compact, occupying 26 KB of memory versus 107 KB for the Mathematica implementation. A user-friendly web interface has also been developed to streamline input handling and computations, making the system more accessible.

\bigskip
\noindent
{\bf Acknowledgments}:
We thank Stefan Meinel for providing the Mathematica code that reads the data file of the QCD lattice form factor. B.S. acknowledges the hospitality of Stefan Meinel at the Department of Physics, University of Arizona, where he gave B.S. a lesson on the mathematics and physics behind the code, as well as the algorithm of the Mathematica code.  This work was financially supported by the Student/Faculty Research Engagement (SFRE) Grant and Summer Undergrad Research Experience (SURE) Grant (B.S.).

\section*{Appendix (A): Code}

The code for analyzing the hadronic transition process is organized into three primary components: constants, data manipulation, and calculations. The constants are defined and adjusted in an initialization phase to enable quick computations. The main function reads file names for form factor parameters, covariances, and higher-order terms, verifies their existence, and opens them if valid. These files must either reside in the same directory as the executable or have their paths specified, including file suffixes. After reading and organizing the data, the files are closed.

Data manipulation functions organize input data into arrays for nominal and higher-order parameters and covariances, aligning them to a specified order for seamless use in subsequent calculations. Once the data is structured, the calculations process it to compute decay rates and their associated errors over a range of data points. Key functions handle tasks such as finding nominal and higher-order form factors, solving the Hamiltonian for decay rates, and calculating errors using numerical differentiation and error propagation.

Finally, the results, including decay rates and errors for each \(q^2\) value, are saved in a file formatted as a data table, simplifying the transition to graphing software for visualization. This approach ensures a streamlined workflow from data input to result generation and presentation.

\pagebreak

%%%%%%%%%%%%%%%%%%%%%%%%%%%%%%%%

\end{document}